\documentclass[a4paper]{article}
\usepackage{graphicx}
\newcommand\TA{{\tt term-abacus}}
\newcommand\TeXmacs{{\TeX{}macs}}
\newcommand\IGNORE[1]{}

\begin{document}

\makeatletter
\@addtoreset{equation}{section}
\makeatother
\renewcommand{\theequation}{\thesection.\arabic{equation}}

\thispagestyle{empty}
\begin{flushright}
AEI-2004-043
\end{flushright}

\begin{center}
{\Large\bf A novel approach to symbolic algebra}
\bigskip\bigskip\bigskip

{\bf T. Fischbacher\medskip\\ }
{\em Max-Planck-Institut f\"ur Gravitationsphysik,\\
     Albert-Einstein-Institut,\\
     M\"uhlenberg 1, D-14476 Potsdam, Germany\\ }
\smallskip
{\small \tt tf@aei.mpg.de}\bigskip

\end{center}

\begin{abstract}
\noindent A prototype for an extensible interactive graphical term manipulation
system is presented that combines pattern matching and
nondeterministic evaluation to provide a convenient framework for
doing tedious algebraic manipulations that so far had to be done manually
in a semi-automatic fashion.
\end{abstract}

\section{Introduction}

\noindent Despite the availability of generic
symbolic term manipulation packages -- computer algebra systems like
Mathematica, Maple, MuPAD, to name just a few well-known ones -- and
despite their wide application in calculation-intensive fields of
study, such as (especially) theoretical physics, the necessity to do
lengthy pen and paper calculations that take days or even weeks still
persists -- especially in string theory and related fields. The
primary underlying reason for this seemingly paradoxical situation
seems to be that the way how calculations are communicated to these
automated systems does not mirror closely enough the way how one
usually thinks about doing a calculation by hand. In particular, while
these systems have their own ideas about implicit canonical reductions
that should be applied to newly generated terms automatically,
communicating a step to them as simple as `do quadratic completion on
the second factor of the third summand' that appears perfectly natural
when mathematicians talk at the blackboard already requires
comparatively sophisticated programming. While the presently available
systems excel at doing lengthy symbol crunching along the lines of
clearly stated procedures, it is especially their inherent weakness in
doing user interaction that makes these systems virtually useless for
the much more explorative calculation style one has to use frequently
e.g. when trying to prove some algebraic property of a lengthy
expression.

In this work, the {\TA} prototype is presented which targets
specifically this problem. While this prototype is still in a too
early stage to do any productive work with it, it both demonstrates
the viability of the underlying approach and has reached a stage in
its development where so many important design decisions have to be
made that further scientific discussion is required first.

\section{The Problem (and its solution)}

To calculate means to apply transformation rules to a term. For
operational purposes, one can regard these as substitutions having two
parts: a {\em pattern} matching part of the structure of a term and a
{\em template} that is parametrized by pieces of the pattern and
dictates how the substitute has to be constructed. These rules can be
quite complicated in some instances, and hence, due to human
imperfection, lengthy pen and paper calculations are inherently prone
to sloppiness errors. These are certainly not the only mistakes one
can make, but the ones that can be avoided most easily by letting a
machine take care of proper application of substitution
rules. Furthermore, the less of one's time -- and more importantly --
mental energy one has to use for purely mechanical tasks, the more one
can concentrate on the important aspects, and hence the deeper one can go.

Using pattern matching to express calculation rules is not a new idea
-- it plays a key role in many of the well known symbolic algebra
systems, such as FORM or Mathematica, where it has proven its
value. While many calculation rules can be formulated conveniently as
pattern matching and substitution rules suitable for a computer (once
an agreement on the underlying term representation has been made), one
fundamental problem is that, quite in general, there often are
different ways how to apply such a rule to a given term in a typical
calculation. Whenever there is a purely mechanical way to decide which
route to follow, the corresponding step in the calculation is not
interesting, as it does not require human intelligence. But evidently,
as every interesting calculation does require human intelligence for
its solution, one is almost bound to encounter the difficulty to find
a way to convey enough information to the machine to completely
specify the particular transformation one has in mind. As this is a
key issue in this work, we want to introduce special terminology here
and from now on call this procedure {\em term clamping}.

It is essential that term clamping has to work in a most unobtrusive
way, requiring as little thought from the end user as possible, or
such a system would be perceived as clumsy and unusable when it comes
to real-world applications.

First and foremost, this means that the amount of extra information
that has to flow has to be minimized -- typing a full command, or
determining values for additional parameters, is already too much of a
hassle. In addition, any information flowing back from the machine to
the user during that process must be presented in such a way that it
does not require any additional interpretation.

On the other side of the coin, it is strongly desirable to find a
clean and concise way to express the logic behind term clamping in the
program code of the implementation of a term manipulation system. The
essence behind term clamping is basically a nondeterministic choice
between different possible futures, introduced into the system by an
intelligent (and hence unpredictable) human user. As humans usually do
calculations on terms which they can easily capture visually as a
whole, the truly minimal amount of information flows between human and
machine if the machine offers a simple visual choice between all
possibilities how to apply a given calculation rule to a term. For the
program, however, this means that we have to produce multiple
solutions to the problem of recognizing a given pattern inside some
larger structure. This is most easily effected by making use of {\em
ambiguous evaluation} \cite{McCarthyAMB,SICP,onlisp}.

Basically, this means that the system which we want to use to
implement such a term manipulation system should support a notion of
making choices and forking a calculation into many different branches,
each of which may produce its own result, or may turn out as futile
and fail to return anything if the corresponding choice is
incompatible with extra requirements we want to impose.

To give a specific example, suppose we wanted to extract all
non-overlapping (unordered) pairs of triples of consecutive
five-letter-words from a sentence such as:
\[
\mbox {\tt The swift small brown horse might never ever allow being shoed}
\]

One would basically want to express an idea like that -- which
admittedly sounds a bit synthetic, but actually has a lot in common in
structure with the calculation rule patterns we are interested in --
in a Scheme \cite{rabbit}\footnote{This work is based on the free
MzScheme implementation, which belongs to the PLT Scheme family
\cite{PLT}} program of roughly the following form\footnote{Clearly,
this has been written with readability for a broad public in mind; a
seasoned programmer would e.g. most probably not let index counting
start at 1. Hint to non-Scheme programmers: Scheme code is read mostly
by indentation, ignoring most of the parentheses.}:

{\small
\begin{verbatim}
(define sentence
  '("The" "swift" "small" "brown" "horse"
    "might" "never" "ever" "allow" "being" "shoed"))

(define empty? null?)

(define (n-th n list)
  (if (empty? list)
      (fail)
      (if (= n 1)
          (first list)
          (n-th (- n 1) (rest list)))))

(define (n-th-rest n list)
  (if (= n 0)
      list
      (if (empty? list)
          (fail)
          (n-th-rest (- n 1) (rest list)))))

(define (find-three-consecutive-5-letter-words list-words)
  (let ((word1 (n-th 1 list-words))
        (word2 (n-th 2 list-words))
        (word3 (n-th 3 list-words)))
    (either 
     (if (and (= (string-length word1) 5)
              (= (string-length word2) 5)
              (= (string-length word3) 5))
         list-words
         (fail))
     (find-three-consecutive-5-letter-words (rest list-words)))))

(define (find-pairs-3x5 list-words)
  (let* ((first-occurrence
          (find-three-consecutive-5-letter-words list-words))
         (second-occurrence
          (find-three-consecutive-5-letter-words
          (n-th-rest 3 first-occurrence))))
    (list (list (n-th 1 first-occurrence)
                (n-th 2 first-occurrence)
                (n-th 3 first-occurrence))
          (list (n-th 1 second-occurrence)
                (n-th 2 second-occurrence)
                (n-th 3 second-occurrence)))))

(all-values (find-pairs-3x5 sentence))

#| Result:

((("swift" "small" "brown") ("horse" "might" "never"))
 (("swift" "small" "brown") ("allow" "being" "shoed"))
 (("small" "brown" "horse") ("allow" "being" "shoed"))
 (("brown" "horse" "might") ("allow" "being" "shoed"))
 (("horse" "might" "never") ("allow" "being" "shoed")))

|#
\end{verbatim}
}

Even if not the details, at least the general structure of this
program should be understandable even for non-Scheme-programmers.
What is especially interesting here is that failure of a calculation
branch can happen at very different places, at different nesting
levels in the calculation, that is, the calculation is highly
non-uniform between different branches.

The Scheme programming language does not provide such highly unusual
constructs like {\tt either} or {\tt fail} or {\tt all-values}. But
remarkably enough, it does provide an universal tool that allows one
to catch the future of any given computation to store it away, or even
call it multiple times (jettisoning the future of the current actual
calculation itself), called {\tt call-with-current-continuation}. With
this, it is possible to seamlessly extend the language by virtually
construct that involves highly nontrivial changes in execution flow --
such as in particular nondeterministic features of exactly the form
presented above -- with little effort. Indeed, it can be done in less
than fifty lines of extra code; this is explained in the appendix.

The example presented here indeed can be regarded as a specific
instance of a matching problem of just the type that covers a large
set of calculation rules. Under the premise of doing algebraic
calculations, our terms generally will be sums of individual summands
that consist of a coefficient plus a series of further factors that
should be treated as noncommuting by default, as we want to be able
to convey extra information in the order of factors. A typical
`local' calculation transformation will then have a form like
\begin{equation}
a_\mu a^\dag_\nu \rightarrow a^\dag_\nu a_\mu + \eta_{\mu\nu}
\end{equation}
where it is understood that $a_\mu a^\dag_\nu$ is a pattern that
matches against a sub-sequence of two operators of types $a a^\dag$
that carry small Greek indices that have to be substituted into the
expression on the right hand side wherever the corresponding actual
indices mentioned in the rule appear.

Besides this, we also want to be able to apply rules where the
location of individual pieces in the sequence of factors does not
matter, and hence matching should succeed regardless of their
position, and especially without first having to move pieces
around. One simple example of such a rule would be:
\begin{equation}
\begin{array}{ll}
\ldots \epsilon_{ijk}\ldots \epsilon_{imn} \ldots \rightarrow&
\phantom+\ldots \delta_{jm}\ldots \delta_{kn} \ldots\\
&-\ldots \delta_{jn}\ldots \delta_{km} \ldots
\end{array}
\end{equation}
One will typically prefer to place the deltas in other places than in
this specific example, but this is not essential here.
What is important is this particular form of a pattern.

There furthermore are rules where one that have both properties at the
same time, that is: one wants to match a number of specific
fixed-length sequences of factors which may appear at various
positions in a term. An example of a rule that is usually expressed in
such a way is the Fierz-Pauli identity, which rather should be
regarded as a collection of various calculation rules that allow to
re-arrange certain expressions that carry four fermions (denoted by
small Greek letters), such as\footnote{See any textbook on quantum
field theory like \cite{Kaku:ym} or \cite{Peskin:ev}}
\begin{equation}
\begin{array}{ll}\displaystyle
\ldots \psi\Gamma^\alpha\phi \ldots \lambda\Gamma_\alpha\eta\ldots\rightarrow&
\phantom{+\frac{4}{4}}\,\ldots \psi\eta \ldots \lambda\phi\ldots\\
&-\frac{1}{2}\,\ldots \psi\Gamma^\alpha\eta \ldots \lambda\Gamma_\alpha\phi\ldots\\
&-\frac{1}{2}\,\ldots \psi\Gamma^{\alpha\beta\gamma}\eta \ldots \lambda\Gamma_{\alpha\beta\gamma}\phi\ldots\\
&-\phantom{\frac{1}{2}}\,\ldots \psi\Gamma^5\eta \ldots \lambda\Gamma^5\phi\ldots
\end{array}
\end{equation}

This is -- up to the issue of ordering pieces -- precisely the
structure of our text matching example: we want to be able to match a
collection of non-overlapping fixed-length subexpressions with
additional constraints in a sequence in all possible ways and let the
user choose. This type of pattern is also so common that we should
coin a special term for it -- let us call this a
sequences-of-factors-pattern, in short {\em sofpa}. At the core of the
{\TA} prototype lies a nondeterministic sofpa-matching engine.

Internally, this matching engine produces a list of sub-sequences
which carry annotations which part of the pattern they matched, or if
they lie between patterns, plus information about identifications of
jokers within these patterns.

While this general structure covers many calculation rules, there are
as well examples of term transformations that cannot be expressed in
such a way. Among those, however, one also finds many re-occurring
structures. One major central concept which re-occurs in many guises
but can not be captured in a sofpa rule is forming a variation of an
entire term (not only a specific summand) along the lines of the
Leibniz rule:
\begin{equation}
\delta(ab) = a(\delta b) + (\delta a) b
\end{equation}

Within the present framework, the approach taken is to first
implement them in a more ad-hoc way, and look for re-occurring
structure that should be abstracted out while the prototype
evolves.

In the present form of the prototype, a sofpa rule is represented
internally as an associative list containing a pattern (which is a
list of chains of factors), a substitution part, and additional
information about highlighting telling which matched parts of a
pattern to display in a visually distinct way. In particular, a rule
like quantum mechanical normal ordering 
\[
a a^+ \rightarrow a^+ a +1
\]
is denoted internally as follows:
{\small
\begin{verbatim}

(define *rule-normal-ordering-a*
  `((pattern . ((,(as-pattern '?a '((a)))
                 ,(as-pattern '?a+ '((a "+"))))))
    (subs .
          #(;; summands
            (1 . 
               #(;; one subs for every factor-block pattern
                 (((a "+")) ((a)))))
            (1 .
               #(()
                 ))))
    (highlighting .
      ((?a . green)
       (?a+ . green)))))

\end{verbatim}
}

To the system, every term is a vector of pairs of a coefficient and a
list of factors, and every factor within a summand is a pair of a
stem and either an exponent or a list of tensor (upper and lower,
which denote contravariant and covariant) indices. The stem itself
consists of a symbol and additional ornaments which are symbol
specific, that is, the system contains a set of hooks where one can
provide arbitrary code that defines the meaning and visual
representation of that particular symbol. Via those means, it is
e.g. possible to extend the system by a definition of a {\tt del}
symbol which carries as ornament some other field plus an index and
renders e.g.  {\tt ((del ((F) (down . mu) (down . nu))) (down . rho))}
visually as $\partial_\rho F_{\mu\nu}$. While one would wish to retain the
highest possible flexibility for extending the system with new
interpretations for term ornaments, there are a few slightly subtle
issues including (but not limited to) behaviour under renaming of
silent indices one has to be aware of. (Upon a closer look, category
theory appears to be a language especially well suited to talk about
these subtleties.)

At a more elementary level, the matcher tries to perform a one-to-one
recursive structural match between pattern and value similar to the
{\tt equal?} Scheme function, but with the additional features that a
special `joker symbol' in the pattern (the default convention -- which
can be changed -- is that all symbols whose name starts with a '{\tt
?}' are joker symbols) matches either an arbitrary value, or if it
occurs more than one time in the pattern, a set of {\tt equal?} 
values. Furthermore, there are other special classes of jokers that
e.g. match a sublist of arbitrary length of a given list (an
inherently nondeterministic specification). One may also place jokers
into the right-hand side substitution template, where they are either
instantiated to the corresponding pieces matched by the template, or
to letters from various alphabets with the additional guarantee that
no such joker is instantiated to a letter that occurs somewhere else
in the summand (this is to conveniently implement the generation of
silent indices). If the pattern contains a function, this function is
called with the corresponding part of the value and information about
previous matches as arguments and may nondeterministically fail, or
provide multiple choices of further successful match information. One
observes that this scheme is flexible enough to easily transfer the
spirit of any complex pattern matching notion to this system, such as
those of guards or as-patterns in Haskell, by using
pattern-matching-function-generating-functions, like {\tt as-pattern},
which maps a joker name and a sub-pattern to a function matching
against the sub-pattern and, if successful, binding the matched value
against the provided joker name.

The design decisions about the internal structure of terms are in part
motivated by the goal to use it for tensor algebra as required for
quantum field theory. While it may seem strange at first to provide
direct support e.g. for such a special detail as tensor indices, which
one might rather like to think of an issue to be resolved at the level
of specifying factor ornaments, this actually turns out to be
necessary. One may catch a glimpse of the underlying issues by
observing that it certainly makes sense to allow factors to carry
powers and provide direct support for this, while powers and indices
are a non-orthogonal concept in the sense that one cannot make sense
of an expression like $\left(C_{abc}\right)^3$.

Clearly, the language in which calculation rules are expressed is
still way too low-level to be of use to end users of the system, but
as implementing application-specific languages (in the broadest sense)
is what systems such as Scheme from the LISP family truly excel
at \cite{SICP}, this is merely a question of experimenting with
different notations until one is discovered that turns out to be
simple, powerful, and well-suited for use by non-schemers.

A further comment has to be made concerning the possibility to use the
flexibility of the system to transfer `sloppy' calculations one-to-one
to the machine -- in the sense that one may choose representations of
factors that are ignorant of certain aspects that are conceptually
vital from the mathematical point of view (such as the dependency of
fields on the particular point in spacetime) but do not matter for
some particular calculation one wants to do. While it is nice to
specify all the mathematical structure in full detail, as this helps
to come to a deep understanding of the subject and discover many
interesting conceptual subtleties\footnote{One may consider especially
the treatise on classical mechanics \cite{SICM} as a prophetical
exemplification of the power of this philosophy}, it is perhaps
nevertheless a good idea not to impose too great restrictions on the
level of rigor to the user, as the ability to leave even conceptually
important details that turn out not to have any influence on the
calculation out of the description does have its advantages.

To summarize this section, the problem of allowing the user to
communicate choices about where to apply a given calculation rule if
there are multiple possibilities is most directly expressed in terms
of nondeterministic evaluation and continuations. This is a suitable
language to formulate rule patterns in a concise fashion, but there
are many details one has to be aware of that require additional
discussion.

\section{Anatomy of the {\TA} prototype}

As mentioned previously, the {\TA} prototype is implemented in
MzScheme, as this provides a lot of highly useful infrastructure such
as lisp-style {\tt defmacro} macros, a lexer and parser as well as
(most important!) continuations and even a framework to implement
continuation-based web services following the ideas presented in
\cite{lwba}.

The first and foremost problem that has to be overcome is to find
means to visually display terms in a convenient way and also allow
user input. The requirement to support user input as well as the
generic problem that solutions built by coercing various independent
components to cooperate which were never intended to do so by making
massive use of interprocess communication typically leads to brittle
systems that may react very badly to version updates of individual
components basically excludes any solution based on employing {\TeX}
to do the rendering. While the idea may be tempting to try to
implement this system as a special Emacs mode, as Emacs at least in
spirit intends to be a substrate for such kinds of application, this
does not work as the text rendering features of Emacs are not
sophisticated enough to do term typesetting at the level required for
this application with it. Furthermore, Emacs Lisp does not support
continuations, not even closures. At first sight, {\TeXmacs}
\cite{TEXMACS} appears as a much more attractive alternative, being an
Emacs-inspired WYSIWYG-style text editor with advanced {\TeX}
rendering capabilities and a proper scheme (FSF's guile \cite{guile}) as
scripting language. Unfortunately, {\TeXmacs} is still quite
power-hungry, and the amount of rendering functionality exported to
Scheme was too small to build such a system on top of it for a long
time (this may have changed by now).

An earlier LISP-based version of the {\TA} prototype used Screamer
\cite{screamer} to implement nondeterminism (which turned out too weak
as it could not handle nondeterministic anonymous functions well and
led to overly clumsy code) and Zebu \cite{zebu} as parser generator,
and employed an own simple renderer that was very loosely inspired by
the way how monadic I/O works in Haskell to implement abstract
rendering functionality which was then used by specialized renderers
to generate {\TeX}, ASCII, as well as graphical output. {\TeX} output
capability is evidently important to be able to directly use results
obtained in {\TA}. ASCII output is important as we obviously need
functionality to use a simple syntax for ASCII term input, and even
with the most powerful system, one might want to do ad-hoc
modifications not covered by any known calculation rule on terms that
are best done by editing an ASCII representation. Graphical output was
implemented by building a tree of typographic glyphs with additional
relative positioning information which were then drawn by help of the
clg \cite{clg} LISP-GTK interface, extended with some own
functions. Eventually -- mostly due to difficulties based on missing
continuation support in LISP -- MzScheme turned out to be a system
much better suited to build this prototype on.

The switch to MzScheme made a very different and quite exciting
approach to the rendering problem possible: as this Scheme
implementation contains a full-fledged webserver and special
infrastructure to build highly interactive continuation-based web
services, the possibility to abuse a web browser with MathML support
as a graphical front-end becomes feasible. (Incidentally, the idea
behind employing continuations for web services is to use them to
implement specialized control flow structures to hide all the
underlying control transfer complexity -- which comes from the web
request-response model in this case. This is quite similar in spirit
to the central idea behind {\TA}.) This is attractive for two reasons:
first, MathML is gradually emerging as a standard for typesetting
formulae that can be used with a variety of different applications,
second, this immediately allows one to provide all the functionality
of this system as a web service. The drawback of such an approach is
that it brings along certain restrictions concerning the user
interface. Basically, experience tells us that in order to use such a
system in a fast and efficient way, one wants to be able to use it via
keystroke commands, which means that JavaScript has to be used to a
certain extent to implement the user side of the system, and
furthermore, there are limitations on the keystroke commands one may
use as some interfere badly with web browser internal keystroke
commands. Another issue is that MathML support is still poor with many
graphical web browsers. At present, the prototype is intended to be a
system which specifically abuses the mozilla firefox
browser\cite{mozilla} as graphical interface which accidentally also
can be used over the internet and not yet a generic browser
independent web application.  Firefox users which have appropriate
fonts for MathML installed\footnote{Debian GNU/Linux users should
install the {\tt latex-xft-fonts} package} can have a peek at an early
stage of the system, which is under active development, at {\tt
http://term-abacus.aei.mpg.de:8000} (the source is also available
there).

The current MathML renderer is a modified variant of the {\TeX}
renderer that was transliterated directly from the LISP predecessor
and still uses string blocks and templates internally. This is bound
to change, as XML (of which MathML is an application) can be embedded
directly into Scheme S-expressions, which is considered a much cleaner
and more powerful approach.

Besides the matching engine, the renderer and the web interface code,
another important component of the system is the term input parser. At
present, the intention behind this parser is to provide the most basic
means to input terms as strings like {\begin{verbatim}-7/2 e**4 X_a_b Q^a^b_alpha + 5 Z_alpha\end{verbatim}}
\noindent only in order to keep things simple, but also be
extensible by allowing users to register (almost) arbitrary extra
parsers for special symbols that process symbol ornaments. The only
restriction on such special user-definable parsers is that symbol
ornaments will be delimited by matching pairs of brackets {$[\,]$};
hence, it is easily possible to introduce e.g. a user-defined parser
for things like lepton spinor factors using a syntax like 
{\tt u[bar,mu;p\_1]} or {\tt v[tau;p\_2]} if the application wants this, but
it is not possible to introduce parsers for ornaments with
non-well-formed bracket structure. At the implementation level, a
two-stage LALR(1) parser is constructed employing MzScheme's parser
generator functionality as it is not possible to let matching brackets
delimit tokens by employing a regular lexer only.

\section{Conclusions and Outlook}

While the idea to build a term manipulation system that is suited for
a much more interactive style of working than all other existing
symbolic algebra packages by using nondeterministic language to
concisely model user interaction nondeterminism at the level of the
implementation is very attractive, and has been shown to be feasible
with very moderate programming effort, this approach still has to
prove its value, as the prototype implemented here is still a
bare-bones system that provides all the abstract functionality to
implement specific term manipulation systems on top of it, but no such
system that uses {\TA} has been constructed yet. At least, preliminary
experiments with an implementation of a thermodynamics-oriented term
algebra of partial derivatives on top of the LISP-based predecessor of
that system seemed quite promising. One interesting smoke test that
should be within reach with justifiable effort would be to implement a
set of transformation rules which allow one to do calculations such as
the derivation of the Lagrangian of eleven-dimensional
supergravity \cite{Cremmer:1978km} as easily as possible. This should
also show where the {\TA} system still requires to be refined and
extended.

\bigskip\bigskip

\noindent {\bf Acknowledgments}

\noindent I want to express my gratitude to Klaus Aehlig for many invaluable
discussions, especially on parser issues.

\bigskip\bigskip

{\noindent \Large\bf Appendix: On nondeterministic evaluation\\
 and Scheme}

While continuation-based techniques are well established in the
functional programming community, they gained surprisingly little
attention (especially when considering their power) in the mainstream
so far. Considering especially their perceived usefulness for building
highly interactive web services -- an approach that was popularized
especially by Paul Graham \cite{lwba} -- they may well be on the verge
of becoming the next hot issue that makes its way into the mainstream
via convoluted paths which has been known to lisp hackers for decades
-- just as it was the case with standardized support for hash tables,
print-read-consistency for textual representations of recursively
structured data \cite{CBCL}, proper garbage collection, and many
others.

The basic idea behind continuations is that there is a symmetry
between calling a function and returning a value from a function, the
latter one being just a `call to the function representing the entire
future of the present calculation'. If we forget about technicalities
such as when to free which type of memory object, then on the
conceptual level, even the return address from a C function on the
stack may just as well be regarded as an extra function pointer
parameter denoting `the entire rest of the program as a callable
function to which we pass on our return value'. With this philosophy
in mind, it is possible to mechanically transform every program to
so-called `continuation passing style' (CPS) where all functions take
as an extra parameter. For a very simple function like a naive
implementation of the factorial, this would look as follows:

{\small
\begin{verbatim}

(define (factorial n)
  (if (= n 0) 1 (* n (factorial (- n 1)))))

;; === the same after a full CPS transform ===

(define (return x)
  (lambda (c) (c x)))

(define (cps= cont-a cont-b cont)
  (cont-a
   (lambda (a)
     (cont-b
      (lambda (b)
        (cont (= a b)))))))

(define (cps* cont-a cont-b cont)
  (cont-a
   (lambda (a)
     (cont-b
      (lambda (b)
        (cont (* a b)))))))

(define (cps- cont-a cont-b cont)
  (cont-a
   (lambda (a)
     (cont-b
      (lambda (b)
        (cont (- a b)))))))

(define (cps-if cont-test cont-then cont-else cont)
  (cont-test
   (lambda (bool)
     (if bool
         (cont-then cont)
         (cont-else cont)))))

(define (cps-factorial cont-n cont)
  (cps-if
   (lambda (c) (cps= cont-n (return 0) c))
   (return 1)
   (lambda (c)
     (cps* cont-n
           (lambda (c)
             (cps-factorial
              (lambda (c) (cps- cont-n (return 1) c)) c))
           c))
   cont))

;; (cps-factorial (return 5) display) ==> 120
\end{verbatim}
}

Note that in the definition of {\tt cps-factorial}, there is not a
single place left where a value is returned; furthermore, execution
order is totally specified now. While transformation to CPS plays an
important role `under the hood' of a scheme system, complex code
written in full CPS style is evidently almost unreadable to human
beings. It needn't be, however, as all continuation-related issues are
hidden from the user, the only exception being just the {\tt
call-with-current-continuation} function (and the values it
generates), which allows the user to get a handle at the future of the
entire program at an arbitrary point to store it away and {\em jump}
back to this place in the program at any point in time, even multiple
times, with all the surrounding context properly set up.

This construct gives us a bewildering flexibility to extend the
language with new control flow constructs. For example,
nondeterministic evaluation as used in the {\TA} prototype may be
implemented along the following lines (the idea being to let {\tt
all-values} jump down deeply into a calculation which contains many
choice points over and over again until all choices have been seen and
the list of all values is passed on to its own continuation):

{\small
\begin{verbatim}
(require (lib "defmacro.ss"))

(define call/cc call-with-current-continuation)
(define __cont-other '())

(define (__all-values lambda-expr)
  (let ((results '()))
    (fluid-let ((__cont-other '()))
      (call/cc
       (lambda (ret) ; catch the continuation of all-values
         (set! __cont-other `(,(lambda () (ret (reverse results)))))
         (set! results (cons (lambda-expr) results))
         ((car __cont-other)))))))

(define-macro (all-values . body)
  `(__all-values (lambda () . ,body)))

(define (choose choices)
  (let ((rest-choices choices))
    (call/cc (lambda (c) (set! __cont-other (cons c __cont-other))))
    (if (null? rest-choices)
        (begin
          (set! __cont-other (cdr __cont-other))
          ((car __cont-other)))
        (let ((next (car rest-choices)))
          (set! rest-choices (cdr rest-choices))
          next))))

;; This must be a macro, since we do not want to eval this and that!
;; primitive method, just along the lines of CHOOSE.

(define-macro (either this that)
  (let ((sym-c (gensym "c-"))
        (sym-todo (gensym "todo-"))
        (sym-next (gensym "next-")))
    `(let ((,sym-todo (list (lambda () ,this) (lambda () ,that))))
       (call/cc
        (lambda (,sym-c) (set! __cont-other (cons ,sym-c __cont-other))))
       (if (null? ,sym-todo)
           (begin
             (set! __cont-other (cdr __cont-other))
             ((car __cont-other)))
           (let ((,sym-next (car ,sym-todo)))
             (set! ,sym-todo (cdr ,sym-todo))
             (,sym-next))))))

(define (fail) (choose '()))

;; (all-values (cons (choose (list 1 (choose (list 2 3))))
;;             (+ 100 (choose (list 10 20 30)))))
;; => ((1 . 110) (1 . 120) (1 . 130)
;;     (2 . 110) (2 . 120) (2 . 130)
;;     (1 . 110) (1 . 120) (1 . 130)
;;     (3 . 110) (3 . 120) (3 . 130))

\end{verbatim}
}

\begingroup\raggedright

\newpage
\includegraphics[height=12cm]{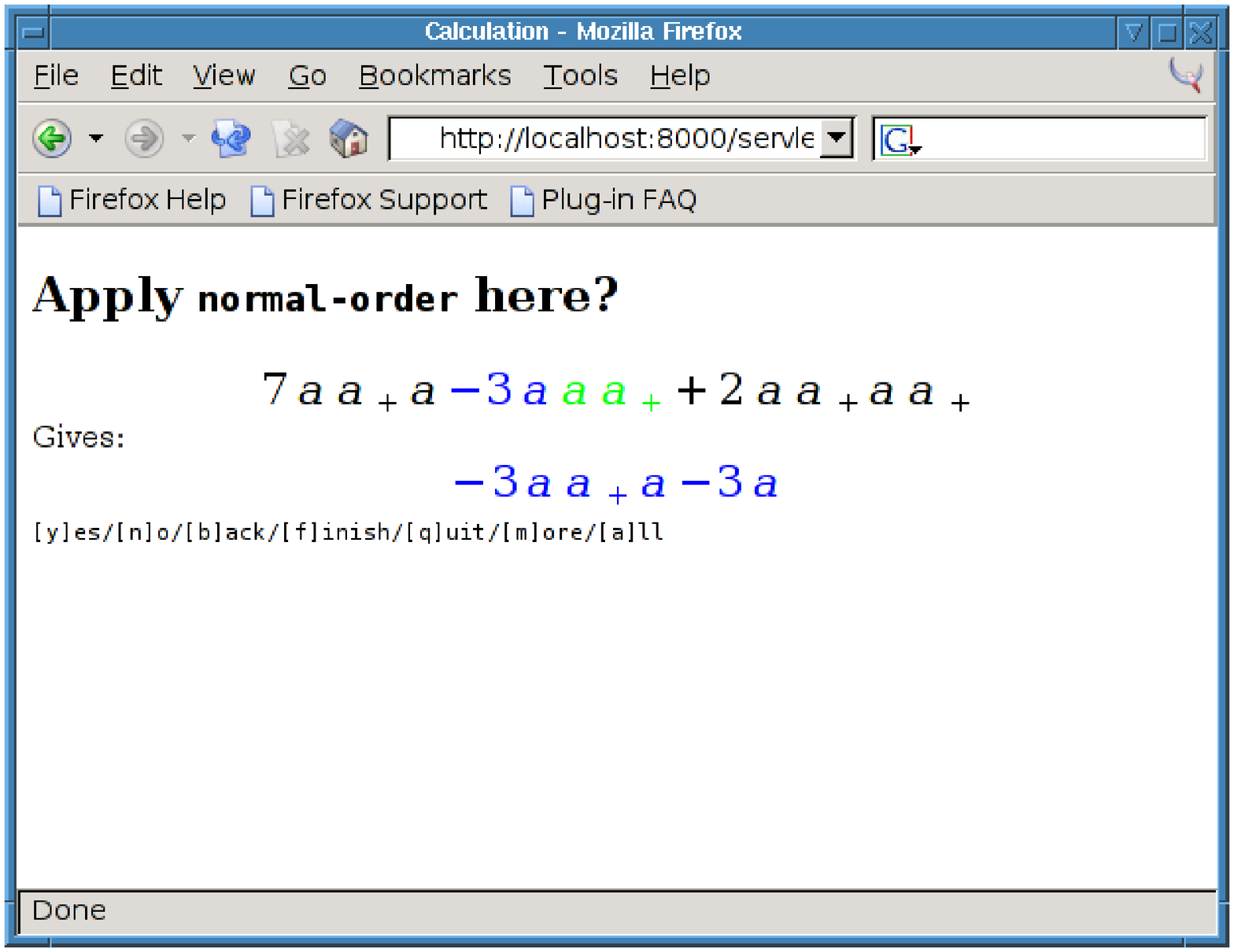}

\end{document}